\documentclass[preprint,12pt]{elsarticle}

\usepackage{graphicx}
\usepackage{amssymb}
\usepackage{lineno}
\usepackage{lipsum}
\makeatletter
\def\ps@pprintTitle{%
 \let\@oddhead\@empty
 \let\@evenhead\@empty
 \def\@oddfoot{}%
 \let\@evenfoot\@oddfoot}
\makeatother

\begin{document}

\begin{frontmatter}


\title{The Bloom Clock}

\author{Lum Ramabaja}

\address{Linz, Austria}

\begin{abstract}
The bloom clock is a space-efficient, probabilistic data structure designed to determine the partial order of events in highly distributed systems. The bloom clock, like the vector clock, can autonomously detect causality violations by comparing its logical timestamps. Unlike the vector clock, the space complexity of the bloom clock does not depend on the number of nodes in a system. Instead it depends on a set of chosen parameters that determine its confidence interval, i.e. false positive rate. To reduce the space complexity from which the vector clock suffers, the bloom clock uses a ''moving window'' in which the partial order of events can be inferred with high confidence. If two clocks are not comparable, the bloom clock can always deduce it, i.e. false negatives are not possible. If two clocks are comparable, the bloom clock can calculate the confidence of that statement, i.e. it can compute the false positive rate between comparable pairs of clocks. By choosing an acceptable threshold for the false positive rate, the bloom clock can properly compare the order of its timestamps, with that of other nodes in a highly accurate and space efficient way.
\end{abstract}


\end{frontmatter}


\section{Introduction}
\label{S:1}

From caching, packet routing and  forwarding, to peer to peer networks, bloom filters are being used extensively in distributed systems. It’s false positive to low space complexity trade-off, can be highly useful for many tasks. In the case of the bloom clock, no new bloom filter variant will be introduced, a lot of smarter people already invented more than enough variants to the originally proposed algorithm \cite{Goodrich2011-vl,Almeida2007-oi,Subramanyam2015-io,Yoon2010-aw,Lall2001-ui,Deng2006-di,Cohen2003-he,Chazelle_undated-dk,Fan2000-qn}. Instead, I will show how bloom filters can be used to generate a partial ordering of events in distributed systems  in a space-efficient way. To some readers the last sentence might sound familiar to the concept of a vector clock \cite{Fidge1987-bm, Mattern1988-dc}. In fact, the bloom clock is an alternative to the vector clock, meant to be used in highly distributed and dynamic systems, where one cannot use a vector clock (due to its space complexity).

Understanding the intuition and theory behind bloom filters and vector clocks is tremendously important  for understanding the bloom clock. This is why in this introduction both these concepts will be briefly described, before continuing to introduce the bloom clock. 

\subsection{The Bloom Filter}
\label{S:1.1}

The bloom filter is a simple, space-efficient probabilistic data structure designed to quickly verify whether an item is present in a set. Its low space complexity comes with a twist: the bloom filter cannot return false negative matches, but it can return false positive matches. In other words, a bloom filter can either tell us that an item is definitely not in a set, or that it might be in a set. 

\begin{figure}[ht]
\centering\includegraphics[width=0.7\linewidth]{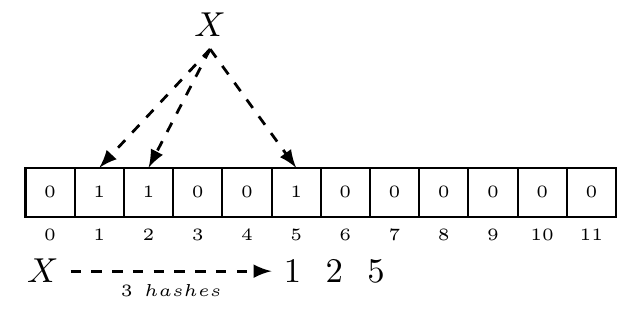}
\caption{Inserting an element into a bloom filter.}
\label{fig:bf}
\end{figure}

Unlike many other data structures, bloom filters do not actually store the items themselves, instead they hash an element $k$ times and then increment the corresponding indices in the bloom filter (where each hash determines which index to be incremented). This introduces false positives, but also results in a data structure with low space complexity, ideal for quick check ups while keeping the memory consumption low. To better understand how this is done, let's go over the algorithm:
\begin{enumerate}
\item Define \textit{k} independent hash functions (where ''\textit{k}'' is the number of hash functions used). The details of how hash functions work, are well beyond the scope of this paper. For now, imagine a hash function as a black box, that maps inputs of arbitrary size to outputs of fixed size.
\item Define a bit array with \textit{m} bits, all set to zero.
\item To add an element to the bloom filter, first hash the element with \textit{k} hash functions (defined at step one). Each hash is used to point to an index of the bit array (defined at step two). Those positions in the bit array are then switched from zero to one, as depicted in figure \ref{fig:bf}.
\end{enumerate}

To check if an element is in the bloom filter, we hash it and lookup the indices. Logically, if one of the \textit{k} indices in the bloom filter is zero, the bloom filter has never seen that element before, otherwise it would not be a zero, but a one. This is why false negatives do not occur in bloom filters. So one can easily prove if an element is not in the array, but what if one wants to prove if an element \textit{is} in the array? This is where false positives might occur.
\begin{figure}[ht]
\centering\includegraphics[width=0.7\linewidth]{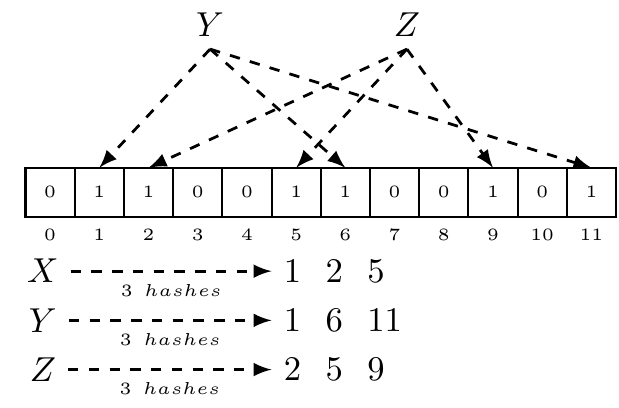}
\caption{ Inserting multiple elements into a bloom filter. ''\textit{X}'' was never inserted, yet it appears as if it has, i.e. a false positive has occurred.}
\label{fig:insertingBf}
\end{figure}

As depicted in figure \ref{fig:insertingBf}, let's assume there are three hash functions. To insert the element $Y$, it is first hashed three times. Let's assume that the resulting hashes are $\{11, 6, 1\}$. Those indices (11, 6, 1) in the bloom filter are then switched to one. The same thing is done to the element $Z$ - it is hashed three times (this time to $\{5,2,9\}$) and then those indices in the array are switched to one. There are now two elements in the bloom filter. If someone checks for $Y$ or $Z$ in the filter, it would appear that they are present, as all the values of those indices are one. But let's say someone checks for element $X$ in the bloom filter. The element happens to get mapped to $\{1,5,2\}$, all of which are already set to one in the bloom filter. There is no way for us to know if $X$ really was inserted into the bloom filter or not. This is also known as a false positive - the bloom filter says that $X$ was inserted even though it was not.

The good news is that one can quite freely control the rate of false positives by tweaking three variables - the size of the bloom filter (\textit{m}), the number of elements inserted into the bloom filter (\textit{n}), and the number of hash functions (\textit{k}) used to hash the elements. The formula for the false positive rate for the bloom filter can be written as: 

\begin{equation}
\label{eq:bf_fpr}
\left(1-\left(1-\frac{1}{m}\right)^{kn}\right)^k
\end{equation}

This formula makes intuitive sense, $1-\frac{1}{m}$ is the probability that a given bit in the bloom filter is not set to one by a certain hash function. $\left(1 - \frac{1}{m}\right)^k$ is the probability that a particular bit is not set to one, given that $k$ hashes can potentially point to it. $\left(1-\frac{1}{m}\right)^{kn}$ is thus the probability that a particular bit is still zero after $n$ elements. And finally, $\left(1-\left(1-\frac{1}{m}\right)^{kn}\right)^k$ gives the probability that $k$ indexes will be 1 after inserting $n$ elements, aka the false positive rate of a bloom filter.

\begin{figure}[ht]
\centering\includegraphics[width=0.7\linewidth]{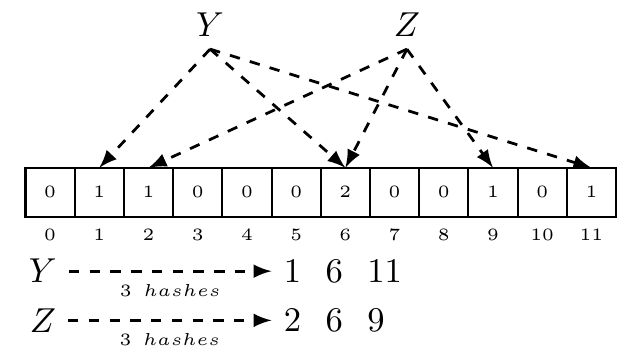}
\caption{Illustration of a counting bloom filter. The counting bloom filter can be incremented to values larger than one.}
\label{fig:cbf}
\end{figure}

This is the basic idea of bloom filters. As I mentioned before, there are many variants of the bloom filter, one of them being the counting bloom filter \cite{Fan2000-qn}. This kind of bloom filter, instead of having only two possible states per index (zero or one), can be incremented to have values larger than one. If two insertions point to the same index, the value at that index increments appropriately, i.e. if the value for an index was zero and two hashes from two elements point to it, it would become two after the insertions, as depicted in figure \ref{fig:cbf}.

I specifically mention the counting bloom filter, because it is the bloom filter variant used in this variant of the bloom clock, and it is what allows us to determine the partial ordering of events between nodes. It is important to note that one does not have to use the counting bloom filter for a bloom clock to work. One might also use other variants, like the d-left counting bloom filter \cite{Bonomi2006-lx}, or the bitwise bloom filter \cite{Lall2001-ui}, or any other kind of bloom filter. Other alternatives might even be more practical in real-world applications, but for the sake of simplicity, I am going to continue using the counting bloom filter when explaining the bloom clock.

\subsection{The Vector Clock}
A vector clock is a simple yet useful algorithm for determining the partial ordering of events in a distributed system. In other words, it allows us to determine the ''passage of time'' in a distributed system, without having to trust physical timestamps. One might wonder why use logical timestamps, instead of the physical timestamp of each node in a distributed system. The reason for that, is that one cannot guarantee that time passes in the same way for all nodes in the network. One node might label an event with a timestamp that already passed for the other nodes, creating causality violations. For this reason, it is crucial for distributed systems to be able to deduce causality without relying on physical timestamps. The vector clock is one such algorithm. There are also other logical clock algorithms, such as the matrix clock \cite{Singh_undated-jz}, or the lamport clock \cite{Lamport1978-is}, on which the vector clock is based on. Before explaining right away how the vector clock determines the partial order of events, let's first define what a partial order is:

In order theory, a set is said to have a total order, if for any element $a$ and $b$, a comparison is possible, i.e. either $a \leq b$ or $a \geq b$. For example: Every element in set $\{5,6,7,8,9,10\}$ is comparable to any other element in that set. We thus can easily know which element happened-before another element. A partially ordered set on the other hand, is a set where only certain pairs of elements are comparable, i.e. one element precedes the other in the ordering, but not every pair of elements is necessarily comparable.

\begin{figure}[ht]
\centering\includegraphics[width=0.4\linewidth]{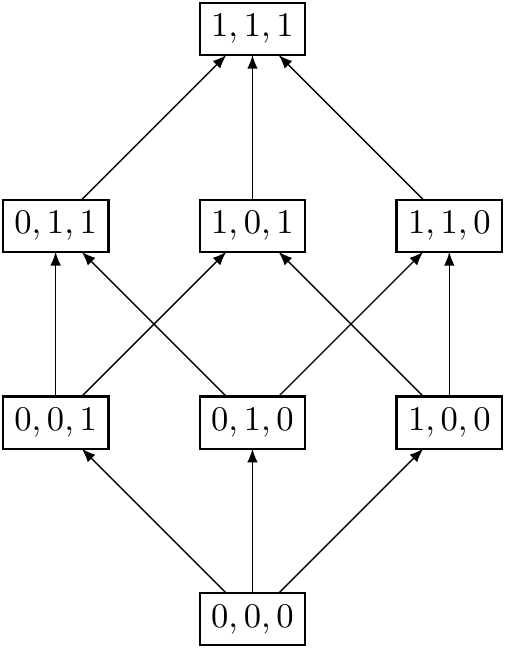}
\caption{A partially ordered diagram, also known as a join-semilattice.}
\label{fig:jointsemilattice}
\end{figure}
As an example, let's look at the diagram in figure \ref{fig:jointsemilattice}. The diagram shows a set $S$ with eight timestamps. We say that an element in $S$ ''happened-before'' another element, if and only if every value of element $a$ is less than or equal to every corresponding value in element $b$. So for example element $(1,0,0)$ happened before element $(1,1,0)$, because none of the values in element $(1,0,0)$ are greater than in element $(1,1,0)$ - We say that $(1,0,0)$ happened-before $(1,1,0)$. If on the other hand we try to compare element $(1,1,0)$ and $(1,0,1)$, one can see that both elements have values larger than the other element at some indices. We say that this pair is \textit{not comparable}. One cannot determine which element occurred before the other one. In the context of the vector clock, incomparable timestamps can occur when nodes send messages concurrently, or when nodes were not able to communicate with one another for some time.

The vector clock algorithm follows the same principles as described above. Instead of relying on a total order to determine causality (e.g. by using synchronized physical timestamps between nodes),  the vector clock captures event dependencies by generating a partial order. Let's now go over the algorithm to see how the vector clock achieves to determine causality:
\begin{enumerate}
\item Let there be four nodes: $A$, $B$, $C$, $D$. Each node initializes a vector of size $N$, where $N$ is the total number of nodes in the system.  All these vectors are initially set to zero. 
\item Each time a node sends an event, it increments its logical clock in its vector by one. As an example, if an internal event occurred in node $A$, its vector changes from $\{A:0, B:0, C:0, D:0\}$ to $\{A:1, B:0, C:0, D:0\}$. After incrementing its value, it broadcasts its new vector to all other nodes.
\item Each time a node receives an event, it increments its own logical clock in the vector by one, and then updates each other value in the vector by taking the maximum value of its vector and the receiving vector. As an example, if $A$'s vector is $\{A:2, B:1, C:3, D:2\}$ and it receives vector  $\{A:2, B:2, C:1, D:2\}$, its new vector would be  $\{A:3, B:2, C:3, D:2\}$. 
\end{enumerate}
It is important not to forget the definition of a partial order: \textit{A partially ordered set is a set where certain pairs of elements are comparable}. Note how in the example in the last bullet point, there is no way to determine the order of the internal vector and the receiving vector, as the two vectors are not comparable (both vectors have values greater than the other one at certain indices). If on the other hand, the internal vector were $\{A:2, B:1, C:3, D:2\}$ and the received vector $\{A:2, B:2, C:3, D:3\}$, we would conclude that the internal vector happened-before the receiving vector, because every value in the receiving vector is greater than or equal to the values in the internal vector. If on the other hand the opposite were to happen, i.e. the internal vector is $\{A:2, B:2, C:3, D:3\}$ and the receiving vector is $\{A:2, B:1, C:3, D:2\}$, one could conclude that the node that sent the vector is not up to date (as the internal vector exceeds the received vector in order).
\begin{figure}[ht]
\centering\includegraphics[width=0.9\linewidth]{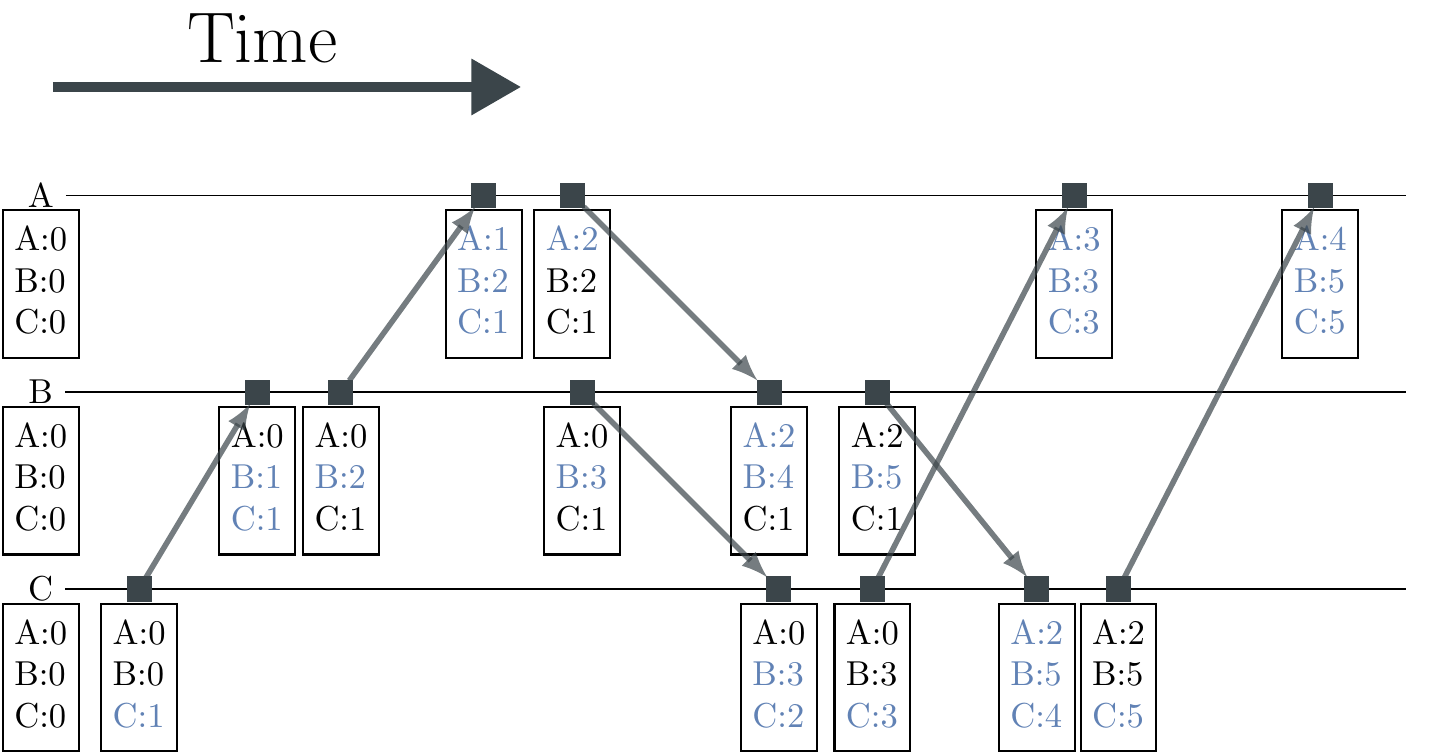}
\caption{An illustration of a vector clock.}
\label{fig:vectorClock}
\end{figure}
The vector clock can be better understood in practice by observing figure \ref{fig:vectorClock}. 

Let's say we take timestamp $\{A:2, B:2, C:1\}$ from node $A$, timestamp $\{A:2, B:4, C:1\}$ from node $B$, and timestamp $\{A:0, B:3, C:2\}$ from node $C$, and we want to compare them. We can see that $B$'s timestamp is comparable and happened-after $A$'s timestamp. Each value of $B$'s timestamp is larger than or equal to $A$'s timestamp. We cannot do the same thing however when comparing $C$'s timestamp to $A$'s timestamp. Both timestamps have some values that are larger than the value in the other timestamp. This means that $A$ and $C$ did not receive the events of the other node, and thus have no information of one-another at that point in time. As a result we cannot conclude which event happened before the other event. Because of this, $A$'s and $C$'s timestamp are not comparable. 

Following these very simple rules, one can determine the partial order of events in distributed systems. This simple yet powerful algorithm is extremely useful for systems that need to know the order of events without relying on local times. But ''with great power, come great costs'' - It turns out that causal ordering is quite expensive \cite{Charron-Bost1991-xm, Schwarz_undated-fl}. The space needed for each vector clock is $O(N)$. This is no problem for small distributed systems, but what about large networks, that might also be dynamic?

\section{Disadvantages of the Vector Clock}
\label{S:2}

Even though the vector clock is a very useful and cleverly designed algorithm, it has its flaws. Because of its space complexity, which is $O(N)$ (where $N$ is the number of nodes in the system), it is difficult to apply it to larger networks. Each time a node sends a message, it has to send its whole vector. For a highly distributed system, this is too costly.
This might also be the reason why vector clocks are mainly used in databases, where the number of nodes is relatively low and static.
In short, the vector clock's dependency on the number of nodes is ultimately the reason why it does not scale well. In highly distributed and dynamic networks, nodes simply cannot communicate by using vectors that take up $O(N)$ space. Furthermore, whenever a node enters or leaves the system, each vector in the system has to get updated appropriately, which is also costly for large networks.
Another issue is the generation of ever larger values with each vector update. If nodes in a distributed system have to send regularly vectors to other nodes, it would be desirable to send vectors with low integer types, to keep a lower transmission size. 
In the next part, I will show how the bloom clock overcomes these fundamental issues, by turning this deterministic problem into a probabilistic problem.

\section{The Bloom Clock}
The bloom clock is a space-efficient, probabilistic data structure used for partial ordering of events in distributed systems. The bloom clock, like the vector clock, can autonomously detect causality violations. Unlike the vector clock however, the size of the bloom clock does not depend on the number of nodes in the system.  Instead it depends on the chosen parameters for its false positive rate. Besides generating a partial order of events, the bloom clock’s timestamps can also be used as a cryptographic proof that an event occurred at a certain point in time. Such a proof however is only possible if the newly assigned timestamp was generated by two previously comparable timestamps. A more detailed example of this is made later in this section. Let's first look at the algorithm behind the bloom clock, and then continue from there:
\begin{enumerate}
\item Define the counting bloom filter's size $m$ and the $k$ hash functions used for the events (remember from section \ref{S:1.1}, one does not have to necessarily use a counting bloom filter, other variants can also work). From this point on, I will refer to the counting bloom filter as simply ''bloom filter'' for convenience.
\item Each time a node has an internal event, it hashes that event with $k$ hash functions and increments its bloom filter. It then sends that bloom filter to all other nodes. As an example, if an internal event occurred in node $A$, its bloom filter might change from $[0,0,0,0,0,0]$ to $[0,1,1,0,0,1]$.
\item Each time a node receives an event, it updates its bloom filter by taking the maximum value of its bloom filter and the receiving bloom filter. As an example, if $A$'s bloom filter is $[0,2,1,0,1,2]$ and it receives the bloom filter $[1,2,2,0,0,2]$, its new bloom filter would be $[1,2,2,0,1,2]$.
\end{enumerate}
Note how in the example in the last bullet point, there is no way to determine the order of the internal bloom filter and the receiving bloom filter, as the two bloom filters are not comparable (both bloom filters have values greater than the other one). The implications of all this are quite interesting. We know that false negatives cannot occur in bloom filters. in the context of the bloom clock, a false negative represents the incomparability of two bloom filters. Bloom filter $A$ cannot be ''inside'' bloom filter $B$, if $A$ has some values larger than $B$. This means that we can always determine if two Bloom clocks are not comparable. 

If bloom filter $A$ on the other hand, has each value smaller or equal to bloom filter $B$, we can deduce that $A$ either precedes $B$, or that it is a false positive. What is meant with a false positive in this case, is that the event that generated bloom filter $A$ might have never happened before $B$, but it happens to be ''inside'' $B$, i.e. $B$ happens to have each value larger than or equal to $A$. In the context of the bloom clock, the false positive rate shows how likely it is for one bloom filter to be overlapped by the other, given the number of increments both arrays have. In other words, a false positive is when we think that there is an order between two bloom filters, but in reality the order is not true.  Assuming that bloom filter increments are random, one can immediately see that the larger the difference: 

\begin{equation}
\label{eq:dif}
\sum_{i=0}^{m}|B_i-A_i|
\end{equation}

between two bloom filters is, the more likely it is for the comparison to be a false positive. We can compute the false positive rate for the bloom clock as:
\begin{equation}
\label{eq:bc_fpr}
\left(1-\left(1-\frac{1}{m}\right)^{\sum_{i=0}^{m}(B_i)}\right)^{\sum_{j=0}^{m}(A_j)}
\end{equation}
Where 

\begin{equation}
\label{eq:bc_comp}
\sum_{i=0}^{m}(B_i) \geq \sum_{j=0}^{m}(A_j)
\end{equation}

The false positive rate in the context of the bloom clock, tells us the confidence interval for knowing if two bloom filters have a real order or not, i.e. \textit{it shows how likely it is for the values of one bloom filter to be overlapped by the values of the other bloom filter, given the number of increments both bloom filters have}. 

This means that there is a certain window of events that can occur between $A$ and $B$ before we cannot know if the order is a false positive or not. One can however overcome this issue, if nodes in a system store the logical timestamps of past events as well. The node whose bloom filter has larger values, can go through its history of timestamps, pick the timestamp with the smallest difference to that of the other node's timestamp, and verify with high confidence the order, or the comparability between the timestamps of the two nodes.

To better understand the intuition of the false positive rate in the bloom clock, let's take take an example: Let's assume that bloom filter $A$ and $B$ are: $A =  [0,2,1,2,0,2]$ and  $B = [2,2,1,2,1,2]$, and let's denote their individual sums as: 
\begin{equation}
\label{eq:bc_sum}
A_{sum}=\sum_{i=0}^{m}(A_i) \quad and \quad B_{sum}=\sum_{i=0}^{m}(B_i)
\end{equation}

We want to know if $A$ truly happened-before $B$ or if it is a false positive. For it to be a potential candidate, first every value of $A$ has to be smaller or equal to the values in $B$ (we say $B$ ''overlaps'' $A$), which in this example is the case. If that would not be the case, we could immediately conclude that these two bloom filters are not comparable. We want to know if there is truly an order between $A$ and $B$ or if it is a false positive. To do this, we first have to think what the probability would be to get a bloom filter that perfectly overlaps $A$, if the way we got that bloom filter was by randomly incrementing an empty bloom filter $B_{sum}$ times. Logically, the larger $B_{sum}$ is compared to $A_{sum}$, the more likely it it is to get at the end a bloom filter that overlaps $A$. The closer $A_{sum}$ is to $B_{sum}$ on the other hand, the smaller the chances are for $B$ to overlap $A$ by accident. 

Formula \ref{eq:bc_fpr} does this for us. It shows how likely it is for the values of one bloom filter to be overlapped by the values of the other bloom filter, given the number of increments both bloom filters have, i.e. it shows the false positive rate for two bloom filters. Let's unpack the formula to better understand it:  $\frac{1}{m}$ is the probability that a given index in the bloom filter gets incremented.   $\left(1-\frac{1}{m}\right)^{\sum_{i=0}^{m}(B_i)}$ is the probability that a particular index is still not incremented after $\sum_{i=0}^{m}(B_i)$ increments. And finally, $\left(1-\left(1-\frac{1}{m}\right)^{\sum_{i=0}^{m}(B_i)}\right)^{\sum_{j=0}^{m}(A_j)}$
 gives the probability that $\sum_{j=0}^{m}(A_j)$ indexes will be incremented after $\sum_{i=0}^{m}(B_i)$ increments.
 
Thus formula \ref{eq:bc_fpr} allows each node to independently compute the false positive rate between logical timestamps. Getting back to the example where $A =  [0,2,1,2,0,2]$ and  $B = [2,2,1,2,1,2]$, if we want to know if $A$ truly precedes $B$, we simply compute $(1-(1-\frac{1}{6})^{10})^{7} = 0.29$. So there is a $0.29$ chance for $A$ not to proceed $B$ in reality. If $B$ has a past timestamp in its history that is closer to $A$, it could use that timestamp to further lower the false positive rate, or to prove the comparability of the timestamps. Bloom filters in real-world applications are of course much larger than in this example, to reduce the chances for hash collisions between events, and to have lower false positive rates.
\begin{figure}[ht]
\centering\includegraphics[width=1.0\linewidth]{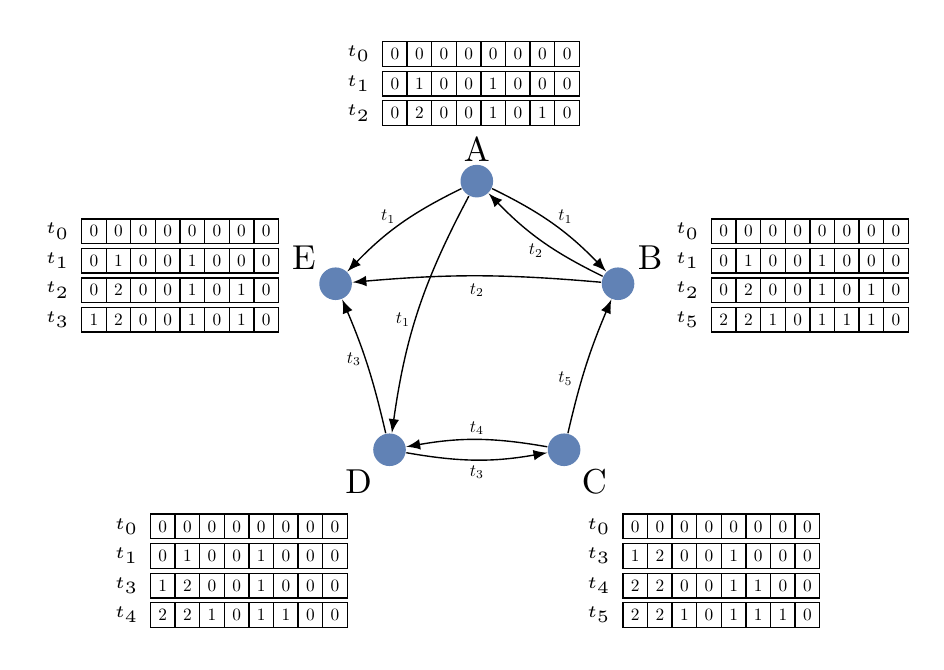}
\caption{A visualization of the bloom clock.}
\label{fig:bloomClockGraph}
\end{figure}

Let's look at figure \ref{fig:bloomClockGraph} for a concrete example of how the bloom clock works in a system. We can see a graph of nodes sending and receiving bloom filters at different points in time. At first the graph might look like a confusing mess, but after breaking it down it becomes quite easy to understand. 

Each node is denoted by an uppercase letter and has the history of its timestamps (bloom filters) next to it as a table. Every message sent to the other nodes has a value $t_i$, where $i$ determines the order in which the events occurred. To keep track of the bloom filter changes in this illustration, I have denoted the corresponding $t_i$ next to each timestamp in each table.

First all nodes are initialized with a zero bloom filter. Node $A$ is the first one to have an event (one can see that because of $t_1$), it hashes the event and updates its bloom filter appropriately (from $[0,0,0,0,0,0,0,0]$ to $[0,1,0,0,1,0,0,0]$). It then sends the new bloom filter to all other nodes. For some reason node $C$ did not receive $A$'s message. The nodes $B$,$D$,$E$ on the other hand increment their latest timestamp so that it overlaps with their previous bloom filter and the receiving bloom filter, i.e. they apply a \textit{max} function that takes the largest value of both bloom filters. Since their previous bloom filter at this point was the zero bloom filter, the new bloom filter for $B$,$D$,$E$ will simply be the bloom filter they received form $A$ (denoted by $t_1$ in the example). The second message comes from node $B$, it updates its latest bloom filter, and sends it to the other nodes. In this example, $C$ and $D$ did not receive the message. As we can see, up to $t_2$ the tables for $A$,$B$,$E$ are the same, meaning no concurrent (incomparable) events happened so far. In fact, if someone were to send together with the timestamps, the event data as well, $A$,$B$, and $E$ would be able to cryptographically prove that they received a given event at that moment. Any of the nodes $A$,$B$,$E$ could take their bloom filter at $t_1$ deduce it from their bloom filter at $t_2$ and get the hashes of the event at $t_2$. Looking at it the other way around, one can determine if a timestamp in their history was generated by two incomparable timestamps, if deducing a given timestamp with the previous timestamp does not give the hashes of the corresponding event. This works of course only if the nodes communicate the event data as well as the bloom filters with one another. 

Getting back to the example, node $D$ is the next one to send an update to the other nodes. It updates its bloom filter which it received at $t_1$ and sends the new bloom filter this time to $E$ and $C$. Here we have our first incomparable example, when $D$ sends its bloom filter to $E$ ($E$ is still at $t_2$), one can see that the two bloom filters are not comparable ($[0,2,0,0,1,0,1,0]$ and $[1,2,0,0,1,0,0,0]$), i.e. both bloom filters have values larger than the other one. As a result the generated bloom filter at $E$ will be the max of $E$’s previous bloom filter and one received from $D$. The other steps in the graph do not show anything new, and shall be skipped. Following the rules that were already provided, we can relatively easily deduce the next steps.

\section{A comparison between the vector clock and bloom clock}
If a system has a small number of nodes, it makes sense to use a vector clock, as the vector of each node would be smaller in size than the counting bloom filter of each node. Things change drastically however when we compare both clocks in highly distributed systems that are also dynamic (where nodes can enter and leave the network). For such systems, the vector clock is simply not suitable. While the vector clock increases linearly in size for each node, the size of the bloom clock stays constant. While the nodes of the vector clock have to update their vectors to take into account leaving / entering nodes, the bloom clock does not have to change anything. Determining causality in vector clocks is after all very expensive \cite{Charron-Bost1991-xm, Schwarz_undated-fl}. 

Another drawback of the vector clock, is its accumulation of ever larger integers, i.e. its vector needs large integer types. In the case of the bloom clock, once every index is incremented (which on average occurs every $\frac{m}{k}$ events), one can increment a single value outside the bloom filter to keep the values in the bloom filter low. Example: Instead of storing bloom filter $[4,3,3,5,7,4,3,3,5]$, one could store it as $(3)[1,0,0,2,4,1,0,0,2]$. The information in a bloom clock is thus much easier to represent in a smaller integer type. One could in fact do the same thing in the case of vector clocks, but the rate of increments there is fundamentally different from that of the bloom clock. If there are nodes that rarely get to interact, the values of those nodes in the vectors will also rarely increment. Therefore it is much less meaningful to apply this technique in the case of the vector clock. The bloom clock on the other hand can apply this technique much more often, as its data structure does not depend on the number of nodes in the system.

\section{Conclusion}
By introducing the possibility for false positives and a window in which the order of events can be inferred with high confidence, the bloom clock tries to solve the main issue of distributed clocks - their space complexity. As inferring causality is an expensive task in distributed systems, turning the problem of partial ordering into a probabilistic problem can be of high interest for certain systems. The bloom clock appears as a promising data structure for systems such as Conflict-free replicated data types (CRDTs), peer-to-peer networks, etc. 

\section{Acknowledgments}
I would like to express my very great appreciation to Arber Avdullahu for implementing the bloom clock, analyzing performances, and providing valuable feedback which has greatly improved this paper, as well as thank Taulant Ramabaja who introduced me to the concept of vector clocks and their application in the real-world.

\section{References}

\bibliographystyle{model1-num-names}

\end{document}